\documentclass[preprint]{revtex4-1}
\usepackage{graphicx}
\usepackage{epstopdf, epsfig}
\usepackage{amsmath, amssymb}
\usepackage{bm}
\usepackage{here}
\usepackage{graphicx,siunitx}
\usepackage{listings,color,courier,natbib}
\graphicspath{{}}
\usepackage[abs]{overpic}

\newcommand{\kf}[1]{{{\color{black}#1}}}

\draft % marks overfull lines with a black rule on the right

\begin{document}

% Use the \preprint command to place your local institutional report number 
% on the title page in preprint mode.
% Multiple \preprint commands are allowed.
%\preprint{}

\title{Synthetic turbulent inflow  generator using machine learning} %Title of paper

% repeat the \author .. \affiliation  etc. as needed
% \email, \thanks, \homepage, \altaffiliation all apply to the current author.
% Explanatory text should go in the []'s, 
% actual e-mail address or url should go in the {}'s for \email and \homepage.
% Please use the appropriate macro for the type of information

% \affiliation command applies to all authors since the last \affiliation command. 
% The \affiliation command should follow the other information.

\author{Kai Fukami}
%\email[]{kai.fukami@keio.jp}
%\homepage[]{Your web page}
\thanks{{Currently at: Department of Mechanical and Aerospace Engineering, University of California, Los Angeles,
CA 90095, USA}}
%\altaffiliation{}
\affiliation{Department of Mechanical Engineering, Keio University, Yokohama 223-8522, Japan}

\author{Yusuke Nabae}
\affiliation{Department of Mechanical Engineering, Keio University, Yokohama 223-8522, Japan}

\author{Ken Kawai}
%\email[]{kawai-ken@keio.jp}
\affiliation{Department of Mechanical Engineering, Keio University, Yokohama 223-8522, Japan}

\author{Koji Fukagata}
\email[]{fukagata@mech.keio.ac.jp}
\affiliation{Department of Mechanical Engineering, Keio University, Yokohama 223-8522, Japan}
% Collaboration name, if desired (requires use of superscriptaddress option in \documentclass). 
% \noaffiliation is required (may also be used with the \author command).
%\collaboration{}
%\noaffiliation

\date{\today. Published in: {\it Phys. Rev. Fluids} {\bf 4}, 064603 (2019).}

\begin{abstract}
We propose a methodology for generating time-dependent turbulent inflow data with the aid of machine learning {(ML)}, which has a possibility to replace conventional driver simulations or synthetic turbulent inflow generators.  As for the {ML} model, we use {an auto-encoder type convolutional neural network (CNN) with a multi-layer perceptron (MLP)}.
For the test case, we study a fully-developed turbulent channel flow at the friction Reynolds number of ${\rm Re}_{\tau} = 180$ for easiness of assessment.
The {ML} models are trained using a time series of instantaneous velocity fields in a single cross-section obtained by direct numerical simulation (DNS) so as to output the cross-sectional velocity field at a specified future time instant.
From the {\it a priori} test in which the output from the trained {ML} model are recycled to the input, the spatio-temporal evolution of cross-sectional structure is found to be {reasonably well} reproduced by the proposed method. 
The turbulence statistics obtained in the {\it a priori} test are also, in general, in {reasonable} agreement with the DNS data{, although some deviation in the flow rate was found.
It is also found that the present machine-learned inflow generator is free from the spurious periodicity, unlike the conventional driver DNS in a periodic domain.}
As an {\it a posteriori} test, we perform DNS of inflow-outflow turbulent channel flow with the trained {ML} model used as a machine-learned turbulent inflow generator {(MLTG)} at the inlet. 
It is shown that the {present MLTG} can maintain the turbulent channel flow for a long time period sufficient to accumulate turbulent statistics, with much lower computational cost than the corresponding driver simulation. {It is also demonstrated that we can obtain accurate turbulent statistics by properly correcting the deviation in the flow rate.}
\end{abstract}

% insert suggested PACS numbers in braces on next line

\maketitle %\maketitle must follow title, authors, abstract and \pacs
% Body of paper goes here. Use proper sectioning commands. 
% References should be done using the \cite, \ref, and \label commands
%%%%%%%%%%%%%%%%%%%%%%%%%%%%%%%%%%%%%%%%%%%%%%%%%%%%%%%%%%%%%%%%%%%%%%%%%%%
%%%%%%%%%%%%%%%%%%%%%%%%%%%%%%%% CHAPTER 1 %%%%%%%%%%%%%%%%%%%%%%%%%%%%%%%
%%%%%%%%%%%%%%%%%%%%%%%%%%%%%%%%%%%%%%%%%%%%%%%%%%%%%%%%%%%%%%%%%%%%%%%%%%%
\section{Introduction}
To date, various types of inflow generators have been proposed for inflow-outflow simulations of turbulence.  Physically speaking, the most straightforward method is to simulate the natural transition, starting from the laminar velocity profile with superimposed random fluctuations, as was examined by \citet{MMP1993} and also used in a recent direct numerical simulation (DNS) of turbulent boundary layer by \citet{Wu2009}.  Although this method is ideal, it requires high computational cost because of the necessity of a sufficiently long computational domain for laminar-turbulent transition.  Adding relevant fluctuations to the mean velocity profile of already turbulent flow \citep{Smirnov2001} is another option, often called synthetic turbulent inflow generator.  As discussed by \citet{Keating2004}, the synthesized velocity fluctuations should have spectral contents similar to those of actual turbulent flows; otherwise the added fluctuations dissipate quickly.  For this purpose, several attempts have been made to generate random fluctuations having proper spatio-temporal correlations.  \citet{Druault2004} and \citet{Perret2008} reconstructed inflow turbulence from measured experimental data using proper orthogonal decomposition and linear stochastic estimation.  \citet{Klein2003}, \citet{Mare2006}, and \citet{Hoepffner2009} used digital filtering techniques to generate correlated field out of random noise.  Yet another, but seemingly most popular method nowadays is to use an auxiliary (i.e., driver) turbulence simulation with a periodic computational domain.  In order to take into account the spatial development in the periodic driver domain, \citet{Lund1996} proposed a rescaling and recycling of velocity profiles and velocity fluctuations based on the law of the wall, which can be considered as a modified Spalart method \citep{PRS1988}.  As a result, they could successfully reproduce the development of turbulent boundary layer.  Although such a driver-type inflow generators is more straightforward than the sophisticated synthetic turbulence generators introduced above, one of its major drawbacks is its additional computational cost.  {Another, and more crucial drawback is the spurious periodicity issue arising from the streamwise periodicity in the driver simulation, as extensively discussed by \citet{Wu2017}.}

In recent years, machine learning has gathered increasing attentions as a part of the boom on big data and artificial intelligence.  Application of machine learning to fluid mechanics problems has a relatively long history.  For instance, \citet{Lee1997} devised a single layer perceptron, which is a simplest type of neural network (NN), to learn the control input of opposition control \citep{Choi1994} for turbulent friction drag reduction.  \citet{Milano2002} used a multi-layer perceptron (MLP) \citep{Rumelhart1986} to estimate the flow field above the wall from the information on the wall --- it is a surprising fact that they treated more than $26\;000$ inputs for neural network already about 20 years ago.  Owing to the recent active development of machine learning libraries such as TensorFlow and Chainer, machine learning has now become a more handy tool also to fluid mechanics.  Recently, \citet{gamahara2017} attempted regression of the subgrid scale (SGS) stress in large-eddy simulation (LES) using a three-layer perceptron, and succeed in reproducing SGS stresses similarly to those of the conventional Smagorinsky model \cite{Smagorinsky1963,Deadorff1970}.  \citet{Ling2016} performed regression of the anisotropy tensor in Reynolds-averaged Navier-Stokes (RANS) simulations using a specially designed NN with an additional tensor input layer so as to account for the Galilean invariance, and demonstrated a better prediction performance than a simple MLP.  \citet{Huang2017} attempted to predict using NN the difference between Reynolds stresses computed by DNS and RANS in a high Mach number flow.

Among different NN architectures, convolutional neural network (CNN) \citep{LeCun1998} has widely been used in the field of image recognition.  One of the features of CNN is that it can naturally take into account the spatial structure of input data, in contrast to the traditional MLP having a fully-connected neural network architecture.  This feature of CNN is also advantageous when fluid mechanics problems are considered.  \citet{Guo2016} proposed a CNN implementation for real-time prediction of non-uniform steady laminar flow.  Although the result shows lower fidelity than the traditional computational fluid dynamics (CFD), CNN is shown to be able to predict the velocity field faster than the CFD solver.  \citet{Yilmaz2017} applied CNN for prediction of the pressure coefficient on airfoils (note that this is originally a regression problem, but they converted it into a classification problem by making groups of pressure coefficient), and achieved more than $80\%$ test accuracy.  In addition, they mentioned that use of hybrid experimental/computational data as the training datasets has a possibility for better regression performance.  \citet{Zhang2018} proposed multiple CNN structures to predict the lift coefficient of airfoils with different shapes at different Mach numbers, Reynolds numbers, and angles of attack.  One of the useful suggestions from their study is that using an artificial image as the input, which is a colored image corresponding to the input parameters, also improves the test accuracy. 

From these contexts, it would be natural to consider utilizing machine learning to develop a turbulent inflow generator that may replace conventional driver simulations or synthetic turbulent inflow generators.
In the present study, we propose such a turbulent inflow generator based on machine learning.

%%% FIGURE 1 %%%%%%%%%%%%%%%%%%%%%%%%%%%%%%%%%%%%%%%%%%%
\begin{figure}
	\vspace{-10mm}
	\begin{center}
		%\hspace{-30mm}
		\includegraphics[width=1.00\textwidth]{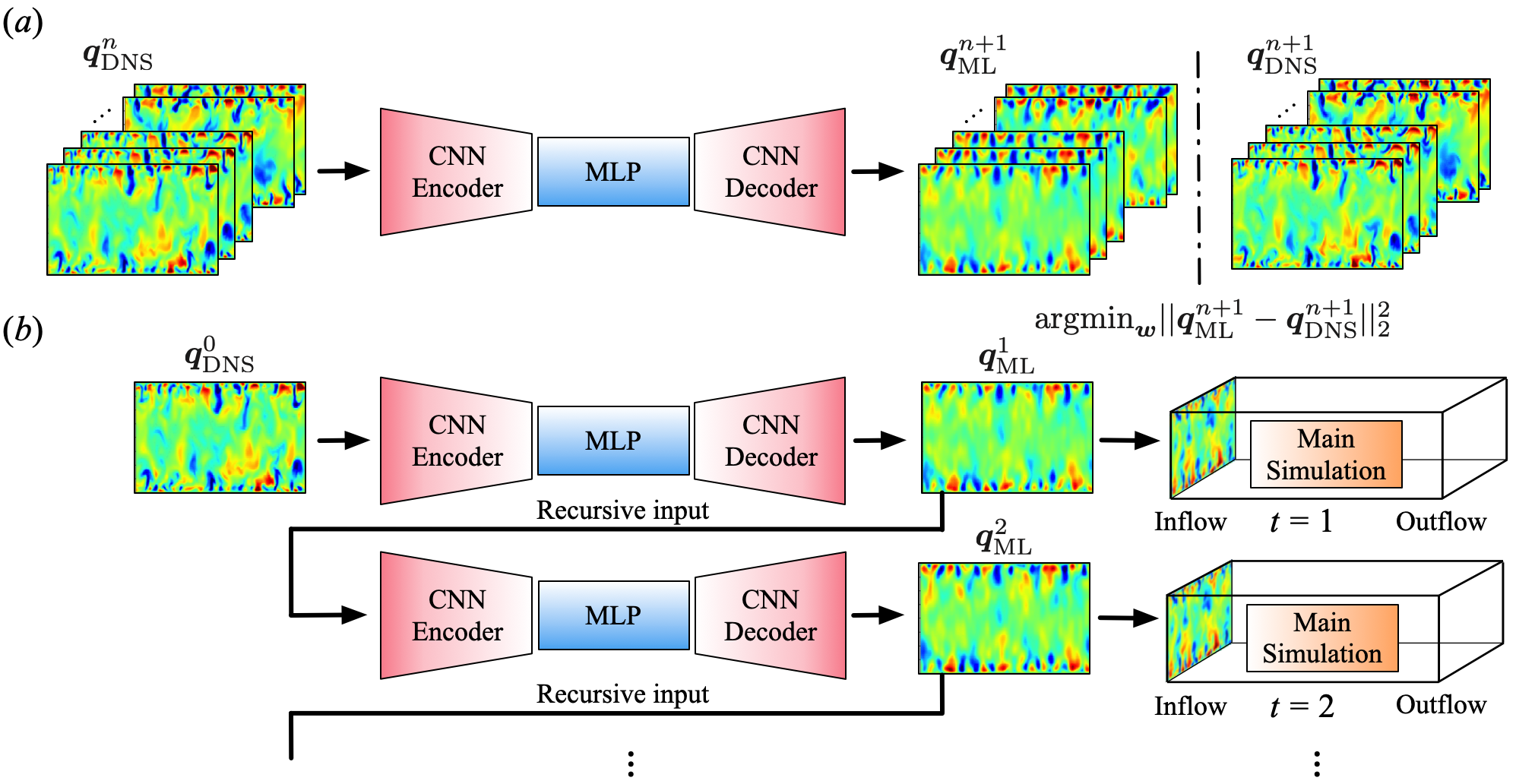}
		\caption{{Schematic diagram of the present machine learning and its use as an inflow generator. ($a$) Training stage; ($b$) {\it a priori} and {\it a posteriori} tests.}}
		\label{fig1}
	\end{center}
\end{figure}
%%%%%%%%%%%%%%%%%%%%%%%%%%%%%%%%%%%%%%%%%%%%%%%%%%%%%%%

{We propose an autoencoder \citep{Hinton2006}-type convolutional neural network (CNN) combined with a multi-layer perceptron (MLP) as a present machine-learning (ML) model.  
As schematically shown in Fig.~\ref{fig1}, the CNN part of the present model works to compress the high-dimensional data of the cross-sectional velocity and pressure field into a lower-dimensional latent space so that important spatial features of the flow are extracted, while the MLP part is used to regress their temporal relationship.
In section~II, the present ML model will be explained in greater detail.}

{The main idea of the present work is illustrated in Fig.~\ref{fig1}($a$).
First, the ML model is trained using the velocity and pressure data in a cross section obtained by direct numerical simulation (DNS) so that the mean squared error (MSE) between the cross-sectional velocity and pressure field at the next time step (i.e., output, $\bm{q}_{\rm ML}^{n+1}$), which is obtained as a response to that at the present time step (i.e., input, $\bm{q}_{\rm DNS}^{n}$), and that of DNS (i.e., the answer, $\bm{q}_{\rm DNS}^{n+1}$) is minimized. 
After the ML model is trained, we perform an {\it a priori test} by recycling the output of the ML model to the input, as indicated by the black arrows with a label ``Recursive input" in Fig.~\ref{fig1}($b$), to investigate whether the spatio-temporal structure similar to turbulence is properly maintained within the ML model.
Note that, at this stage, the DNS data are fed into the ML only once for the initialization (i.e., $\bm{q}_{\rm DNS}^0$ in Fig.~\ref{fig1}($b$)). 
Finally, an {\it a posteriori test} is conducted by inflow-outflow DNS with time-dependent inflow conditions computed using this machine-learned turbulent inflow generator (MLTG).}

%%%%%%%%%%%%%%%%%%%%%%%%%%%%%%%%%%%%%%%%%%%%%%%%%%%%%%%%%%%%%%%%%%%%%%%%%%%
%%%%%%%%%%%%%%%%%%%%%%%%%%%%%%%% CHAPTER 2 %%%%%%%%%%%%%%%%%%%%%%%%%%%%%%%
%%%%%%%%%%%%%%%%%%%%%%%%%%%%%%%%%%%%%%%%%%%%%%%%%%%%%%%%%%%%%%%%%%%%%%%%%%%
\section{Problem formulation and training methods}

\subsection{Training Procedure}
The training datasets are generated using DNS.
For easiness of assessment, we consider a fully-developed incompressible turbulent channel flow as the test case to examine the feasibility of the present approach, although an efficient inflow generator may be appreciated more in simulations of external flows such as spatially developing boundary layers and flows around a body.

The governing equations are the incompressible Navier--Stokes equations,
\begin{eqnarray} 
&\bm{\nabla} \cdot {\bm u} = 0,&\\
&\displaystyle{ \frac{\partial {\bm u}}{\partial t}  =- \bm{\nabla} \cdot ({\bm u \bm u})  -\bm{\nabla} p  + \frac{1}{{\rm Re}_\tau}\nabla^2 {\bm u}}, &
\end{eqnarray}
where $\displaystyle{{\bm u} = [u~v~w]^{\mathrm T}}$ represents the velocity with $u$, $v$ and $w$ being the streamwise ($x$), wall-normal ($y$) and spanwise ($z$) components; $p$ is the pressure, $t$ is the time, and $\displaystyle{{\rm Re}_\tau = u_\tau  \delta/\nu}$ is the friction Reynolds number.  
The quantities are made dimensionless using the channel half-width $\delta$ and the friction velocity $u_\tau$.

The DNS is performed using the finite difference code of \citet{Fukagata2006}, which has been validated by comparison with spectral DNS data of \citet{Moser1999}.
The size of the computational domain and the number of grid points are $(L_{x}, L_{y}, L_{z}) = (4\pi\kf{\delta}, 2\kf{\delta}, 2\pi\kf{\delta})$ and $(N_{x}, N_{y}, N_{z}) = (256, 96, 256)$, respectively.  The grid is uniform in $x$ and $z$ directions, while non-uniform in $y$ direction. 
No-slip boundary condition is imposed on the walls and the periodic boundary condition is applied in $x$ and $z$ directions. 
The DNS is performed under a constant pressure condition at ${\rm Re}_{\tau}=180$.

Time series of velocity and pressure field in a single $y-z$ cross-section computed by this DNS are used as the training data for the {ML} model. 
Since the raw streamwise velocity, $u$, has a strongly skewed distribution due to its mean velocity component, and such an
ill-formed distribution of training data is known to deteriorate prediction using a neural network \citep{Shank1996}, we use the fluctuations, $\displaystyle{{\bm u}' = [u'~v'~w']^{\mathrm T}}$ and $p'$, as the input and output vector, $\bm{q}{\,=[u'~v'~w'~p']}$.
The regression using {the ML} model can be expressed as
\begin{equation}
\bm {q}^{n+1}_{\rm ML}
={\cal {F}}({\bm {q}^{n}_{\rm DNS}}; {\bm {W}})\approx \bm {q}^{n+1}_{\rm DNS},
\label{NN}
\end{equation} 
where $\bm {q}^{n+1}_{\rm ML}$ denotes the cross-sectional field at the next time step computed by {the ML} model, $\bm{q}^{n}_{\rm DNS}$ represents the DNS data at the present time step that are fed as the input to {the ML} model, and $\bm {q}^{n+1}_{\rm DNS}$ is the answer that should be reproduced by {the ML} model.
The nonlinear mapping by {the ML} model is denoted by ${\cal {F}}(\cdot)$, and ${\bm {W}}$ denotes the weights in {the ML} model that are optimized by learning so as to minimize the loss function{, i.e.,} the mean squared error (MSE) between $\bm {q}^{n+1}_{\rm ML}$ and $\bm{q}^{n+1}_{\rm DNS}$.
The time interval between time steps $n$ and $n+1$ in ML is $\Delta t^+ = 1.26$, which corresponds to 10 time steps in the present DNS. 

The code for machine learning has been written in-house by utilizing TensorFlow 1.2.0 and Keras 2.0.5 libraries on Python 3.6.

%%%%%%%%%%%%%%%%%%%%%%%%%%%%%%%%%%%%%%%%%%%%%%%%%%%%%%%%%%%%%%%%%%%%%%%%%%%
%%%%%%%%%%%%%%%%%%%%%%%%%%%%%%% SECTION 2.2 %%%%%%%%%%%%%%%%%%%%%%%%%%%%%%%%
%%%%%%%%%%%%%%%%%%%%%%%%%%%%%%%%%%%%%%%%%%%%%%%%%%%%%%%%%%%%%%%%%%%%%%%%%%%
\subsection{Machine-learned turbulence generator}

%%% FIGURE 2 %%%%%%%%%%%%%%%%%%%%%%%%%%%%%%%%%%%%%%%%%%%
\begin{figure}
	\begin{center}
		\vspace{0mm}
		\hspace{0mm}
		\includegraphics[width=0.70\textwidth]{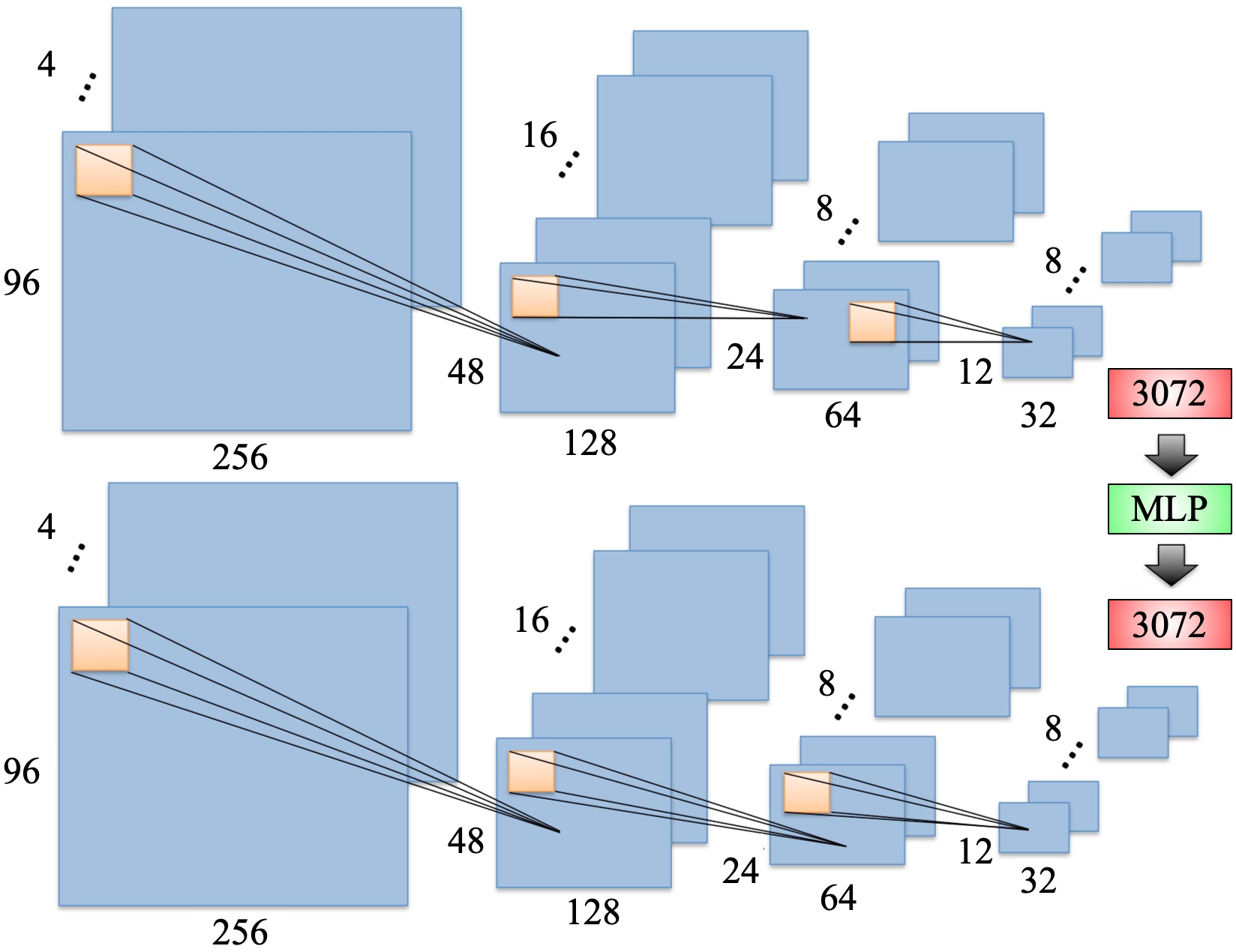}
		\caption{Schematic structure of the machine-learn{ing} model using the auto-encoder type convolutional neural network with multi-layer perceptron (e.g., Case 1).}
		\label{fig2}
	\end{center}
\end{figure}
%%%%%%%%%%%%%%%%%%%%%%%%%%%%%%%%%%%%%%%%%%%%%%%%%%%%%%%

{As shown in Fig.~\ref{fig2},} the basic network structure {of the present ML model} is similar to the standard CNN autoencoder used for image recognition, as can be found, e.g., in Keras tutorial (https://keras.io/).
First, instantaneous $y-z$ cross-sectional data, $\bm{q}^{n}_{\rm DNS}$, are fed into the network.
All input data are standardized so that the mean value is zero and the standard deviation is unity, because it is generally known that if the mean value of input data deviates from zero, the weight update will be affected and the learning speed becomes slower \citep{LeCun2012}.
Since the input data consist of four primitive variables ($\bm u^\prime$ and $p^\prime$) in a single $y-z$ cross-section, the total number of inputs in the present study is $96\times256\times4=98\;304$ per instant{, i.e., the product of the number of computational points in a $y-z$ section and the number of variables}.
This high dimensional input data are first compressed by a sequence of convolution layers.
Then, low-dimensionalized data are passed to fully-connected {multi-layer perceptron (MLP)} layers{, in which the relationship between the low-dimensionalized features at two consecutive time instants are regressed.}
%{These procedures are able to do a global regression in terms of the flow section data and avoid a curse of dimensionality. }
Finally, the data corresponding to the low-dimensionalized field at the next time step are expanded to their original dimension by the deconvolution layers.
The function ${\cal {F}}$ in Eq.~(\ref{NN}) is expressed as
\begin{eqnarray}
{\cal {F}}({\bm {q}^n_{\rm DNS}})={\cal F}_{\rm dec}({\cal F}_{\rm MLP}({\cal F}_{\rm enc}({\bm {q}^n_{\rm DNS}}))), \label{eq6}
\end{eqnarray}
where ${\cal F}_{\rm enc}$, ${\cal F}_{\rm MLP}$, and ${\cal F}_{\rm dec}$ denote the CNN encoder, the MLP layer, and the CNN decoder, respectively.

More detailed structure is shown in Table \ref{table1}.
Conv2D is a convolution layer, in which the data are convolved with filters.
Pooling has the role to compress the data.
Two widely used pooling techniques are MaxPooling and AveragePooling:
the former selects the maximum value, while the latter selects the average value.
We have attempted both pooling models and confirmed that AveragePooling model shows more better accuracy than MaxPooling model in {both training and test processes}; therefore, we {adopt} AveragePooling in the present study.
Upsampling is an operation for a dimension extension, in which the value is copied to the extended dimensions.
In all layers, the filter size is set at $3\times3$ and the pooling size is $2\times2$ for both compression and extension.
As for the activation function, we have tested three different activation functions: 
the hyperbolic tangent (tanh), the rectified linear unit (ReLU), and the sigmoid function.
From this preliminary test, it has turned out that the hyperbolic tangent (tanh) gives the best result {in both training and test processes} in the present study.
The Adam (adaptive moment estimation) optimizer \citep{Kingma2014} is used to optimize the weighting of the {ML} model.

%%%% TABLE 1 %%%%%%%%%%%%%%%%%%%%%%%%%%%%%%%%%%%%%%%%%%%%%%%%%%%%%
\begin{table}[t!]
\begin{center}
\caption{{Detailed structure of the machine-learned turbulence generator (e.g., Case 1)}}
\def~{\hphantom{0}}
\begin{tabular}{ccc}
    \hline \hline 
    Network & Data size & Activation function \\ \hline
    Input &(96,256,4) & - \\
    1st Conv2D&(96,256,16)& tanh\\
    2nd Conv2D&(96,256,16)& tanh \\
    1st AveragePooling 2D&(48,128,16)& - \\
    3rd Conv2D&(48,128,8)& tanh \\
    4th Conv2D&(48,128,8)& tanh \\
    2nd AveragePooling 2D&(24,64,8)& - \\
    5th Conv2D&(24,64,8)& tanh \\
    6th Conv2D&(24,64,8)& tanh \\
    3rd AveragePooling 2D&(12,32,8)& - \\
    1st Reshape&(1,3072)&-\\
    1st {MLP}&(3072)&tanh\\
    2nd {MLP} &(3072)&tanh\\
    2nd Reshape&(12,32,8)&-\\
    7th Conv2D&(12,32,8)& tanh \\
    8th Conv2D&(12,32,8)& tanh \\
    1st Upsampling 2D&(24,64,8)& - \\
    9th Conv2D&(24,64,8)& tanh \\
    10th Conv2D&(24,64,8)& tanh \\
    2nd Upsampling 2D&(24,64,8)& - \\
    11th Conv2D&(48,128,16)& tanh \\
    12th Conv2D&(48,128,16)& tanh \\
    3rd Upsampling 2D&(96,256,16)& - \\
    Output/13th Conv2D&(96,256,4)& - \\
    \hline \hline
\end{tabular}
  \label{table1}
\end{center}
\end{table}
%%%%%%%%%%%%%%%%%%%%%%%%%%%%%%%%%%%%%%%%%%%%%%%%%%%%%%%%%%%%%%%%%%%%%%%

%%% TABLE 2 %%%%%%%%%%%%%%%%%%%%%%%%%%%%%%%%%%%%%%%%%%%%%%%%%%%%%
\begin{table}
\begin{center}
\caption{{Parameters of the machine-learning models, the number of epochs before early stopping, and the resultant mean squared errors (MSE).}}
\def~{\hphantom{0}}
\begin{tabular}{ccccccccc}
    \hline \hline
    Case  & & \# of MLP layers & &Structure of MLP layers & & \# of Epochs && MSE \\ \hline
    Case 1 & &2& &3072--3072& &148&&0.0289 \\
    Case 2 & &2& &192--192& &104&&0.6981 \\
    Case 3 & &3& &3072--3072--3072&&43&&0.6066 \\
    Case 4 & &3& &3072--768--3072&&73&&0.1259 \\
    \hline \hline
\end{tabular}
  \label{table2}
\end{center}
\end{table}
%%%%%%%%%%%%%%%%%%%%%%%%%%%%%%%%%%%%%%%%%%%%%%%%%%%%%%%%%%%%%%%%%%%%%

%%% FIGURE 3 %%%%%%%%%%%%%%%%%%%%%%%%%%%%%%%%%%%%%%%%%%%
\begin{figure}
	\begin{center}
		\vspace{0mm}
		\hspace{0mm}
		\includegraphics[width=0.80\textwidth]{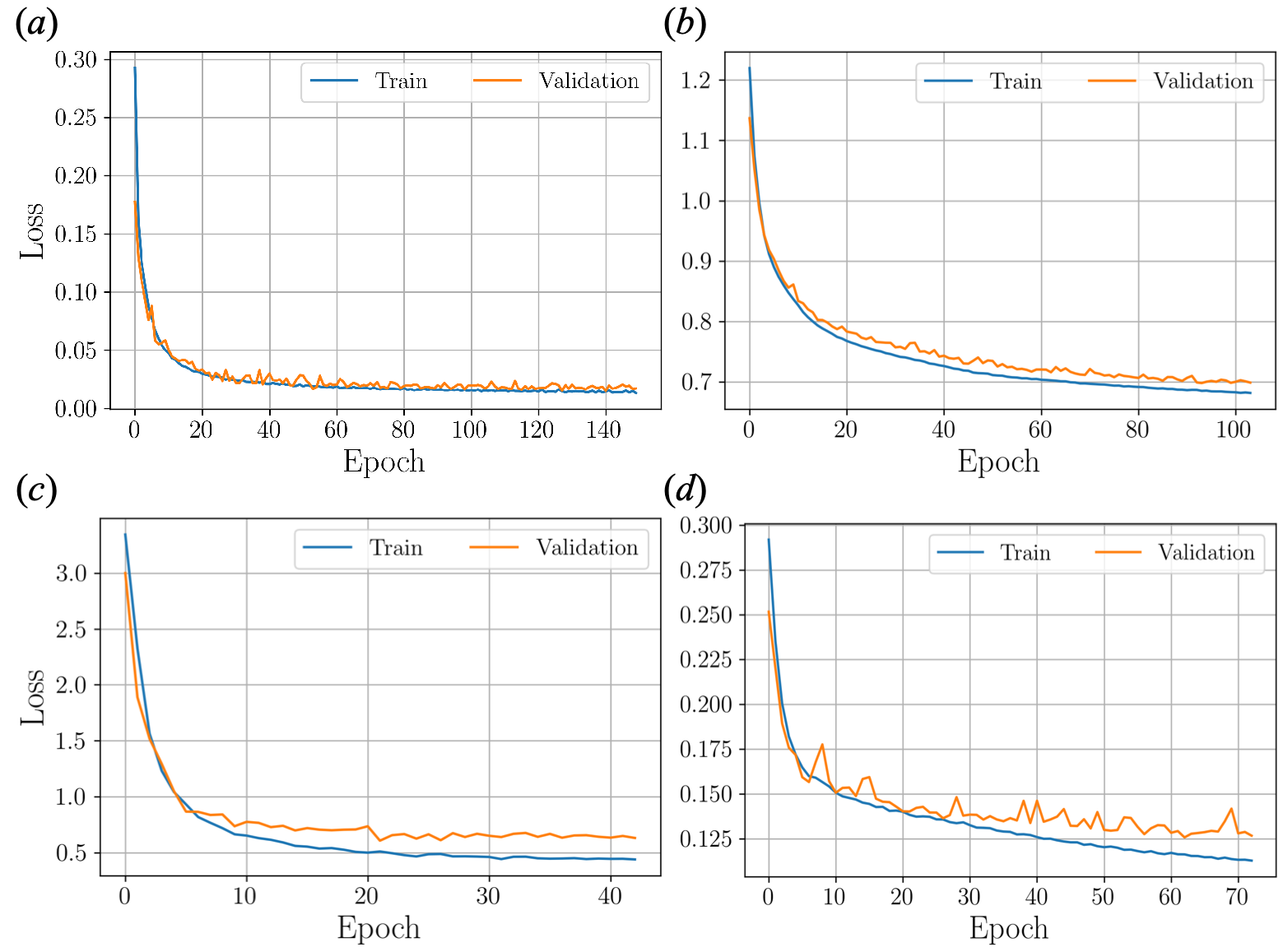}
	    \caption{Learning curve: ($a$) Case 1, ($b$) Case 2, ($c$) Case 3, ($d$) Case 4.}
		\label{fig3}
	\end{center}
\end{figure}
%%%%%%%%%%%%%%%%%%%%%%%%%%%%%%%%%%%%%%%%%%%%%%%%%%%%%%%

{In terms of the computational cost, it is preferable to use as small number of MLP layers and latent data size as possible. 
Simplifying the MLP layer is preferable also for a physical interpretation of the latent space, if possible.
However, oversimplification of the network structure may lead to an insufficient ability to express the essential dynamics.  Therefore, we have examined four cases of machine-learned models by varying the numbers of MLP layers and the latent data size fed to the MLP layer, as shown in Table \ref{table2}, to see their influence on the results.  Case 1 is the base case with two MLP layers whose latent data size is 3072 --- in fact, this was the largest size we could handle under our computer environment. 
In Case 2, a smaller latent data size (i.e., a higher compression ratio) is considered by adding a pair of convolution and deconvolution layers around the MLP layers to the network shown in Fig.~\ref{fig2} and Table \ref{table1}.
Case 3 and Case 4 include an additional hidden layer with the size of 3072 and 768, respectively.}

{In all cases, 10\;000 pairs of snapshots spanning in 12\;600 wall unit time computed by DNS are used.
Among them, $70\%$ is used as the training data, and $30\%$ is used as the validation data.
Note that the pairs of DNS data at two consecutive time instants are fed into the network in a ramdom sequence.
Namely, what the present ML model learns is not the long time series of DNS data themselves but the Navier-Stokes equation distretized with a relatively large time step of $\Delta t^+ \sim 1$.}
Overfitting, where the error for the training dataset is lower than that for the validation dataset, is avoided by employing the early stopping criterion \citep{Prechelt1998} in this study.
A series of continuous 20 epochs is used for the criterion of early stopping. 
The number of epochs trained until this early stopping and the resultant values of MSE are shown in Table \ref{table1}. 

{As can be noticed from the learning curves presented in Fig.~\ref{fig3} and the resultant MSE in Table~\ref{table1}, the learning is most successful in Case 1. 
In Case 2 with higher compression ratio, the resultant MSE is much larger than that in Case 1.
When an extra hidden layer is added (i.e., Cases 3 and 4), the network seems to suffer from overfitting at smaller number of epochs due to the higher degree of freedom (i.e., many parameters) in the latent space.
}

\section{Results and Discussion}
\subsection{$\bm{A\;priori}$ test: recycling within machine-learned model}
As an {\it a priori} test, we recycle the output of the trained machine-learned model to its input for multiple times, with an initial condition taken from a single snapshot of DNS data; namely,
\begin{equation}
\bm {q}^{n+1}_{\rm ML}
={\cal {F}}(\bm {q}^{n}_{\rm ML}; {\bm {W}}),
\end{equation} 
with the initial condition,
\begin{equation}
{\bm {q}^{0}_{\rm ML}}=\bm {q}^{0}_{\rm DNS}.
\end{equation}
The statistics presented below are those accumulated for {10\;000 time steps (i.e., 12\;600 wall unit time)} of this recycling.
Note that we intend to reduce the computation time as compared to traditional driver-type turbulent inflow generators.
Therefore, {the primary purpose of this {\it a priori} test is to see whether} the machine-learned turbulence generator (MLTG) {can} generate self-sustaining inflow turbulence without feeding additional DNS data.

%%% FIGURE 4 %%%%%%%%%%%%%%%%%%%%%%%%%%%%%%%%%%%%%%%%%%%
\begin{figure}
	\begin{center}
		\vspace{0mm}
		\hspace{0mm}
		\includegraphics[width=1.00\textwidth]{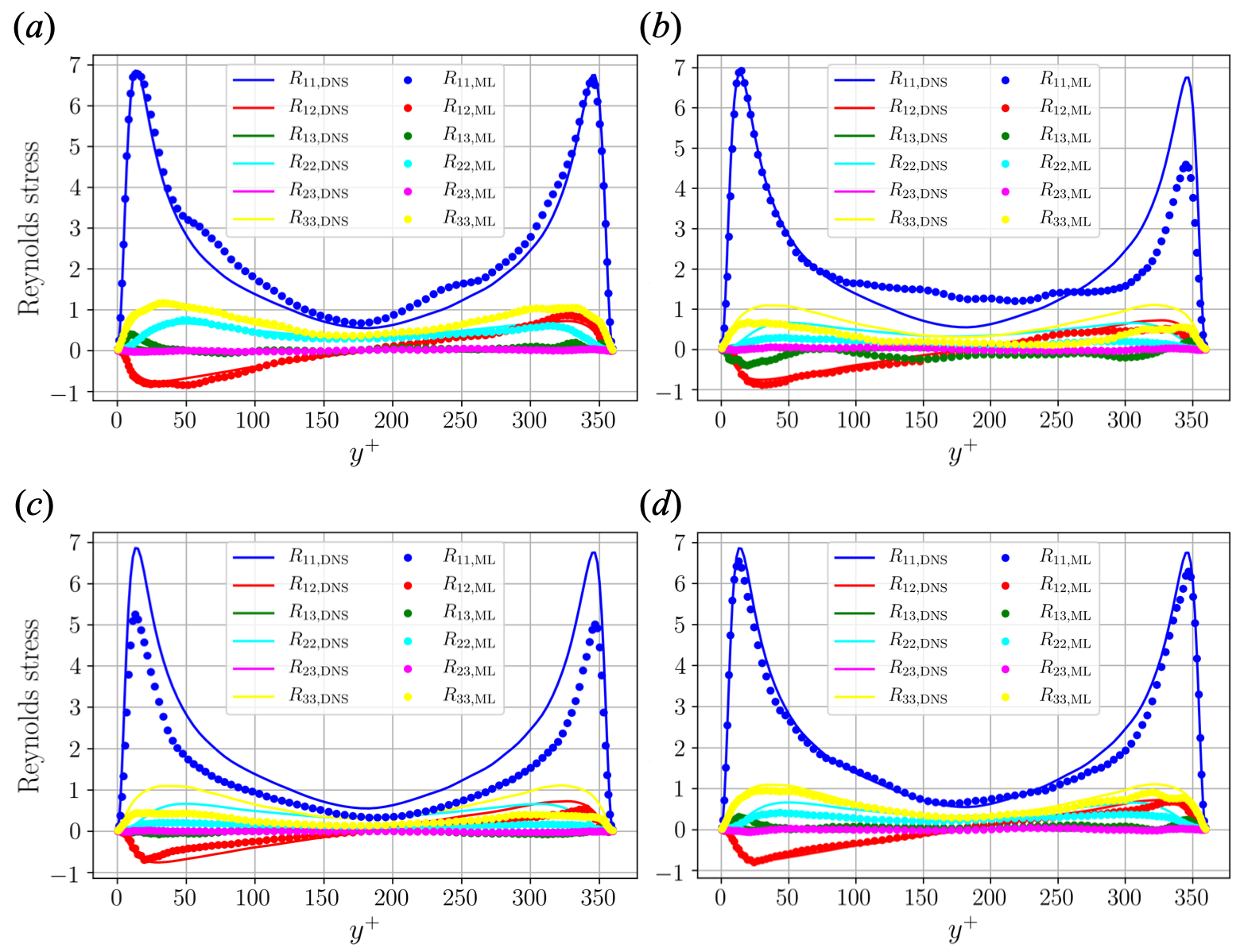}
		\caption{{``Reynolds stresses'' computed based on the raw output, $R_{ij}=\overline{u_i^{\prime+}u_j^{\prime+}}$: ($a$) Case 1, ($b$) Case 2, ($c$) Case 3, ($d$) Case 4.}}
		\label{fig4}
	\end{center}
\end{figure}
%%%%%%%%%%%%%%%%%%%%%%%%%%%%%%%%%%%%%%%%%%%%%%%%%%%%%%%

{
Figure \ref{fig4} shows the ``Reynolds stress'' components computed using the output ``velocity fluctuations'', i.e., $R_{ij}=\overline{u^{\prime+}_i u^{\prime+}_j}$.
We express here the ``Reynolds stress'' and ``velocity fluctuations'' with quotations because the zero-mean properties of $u^\prime_i$ are lost in the present ML model, as discussed later.
Case 1 shows reasonable agreement with the DNS data for all ``Reynolds stress'' components.
Case 2 underestimates the turbulence statistics, especially regarding the components concerning $u$ and $w$, due to overcompression of the latent vector fed to the MLP layer. 
These observations suggest that the data size of the latent space is an important parameter to maintain the physical features. 
In Cases 3 and 4, the computed statistics are qualitatively similar to the reference DNS data; 
however, the accuracy is poor for $R_{11}$ and $R_{33}$ in Case 3 and $R_{22}$ in Case 4.
}

%%% FIGURE 5 %%%%%%%%%%%%%%%%%%%%%%%%%%%%%%%%%%%%%%%%%%%
\begin{figure}
	\begin{center}
		\vspace{0mm}
		\hspace{0mm}
		\includegraphics[width=1.00\textwidth]{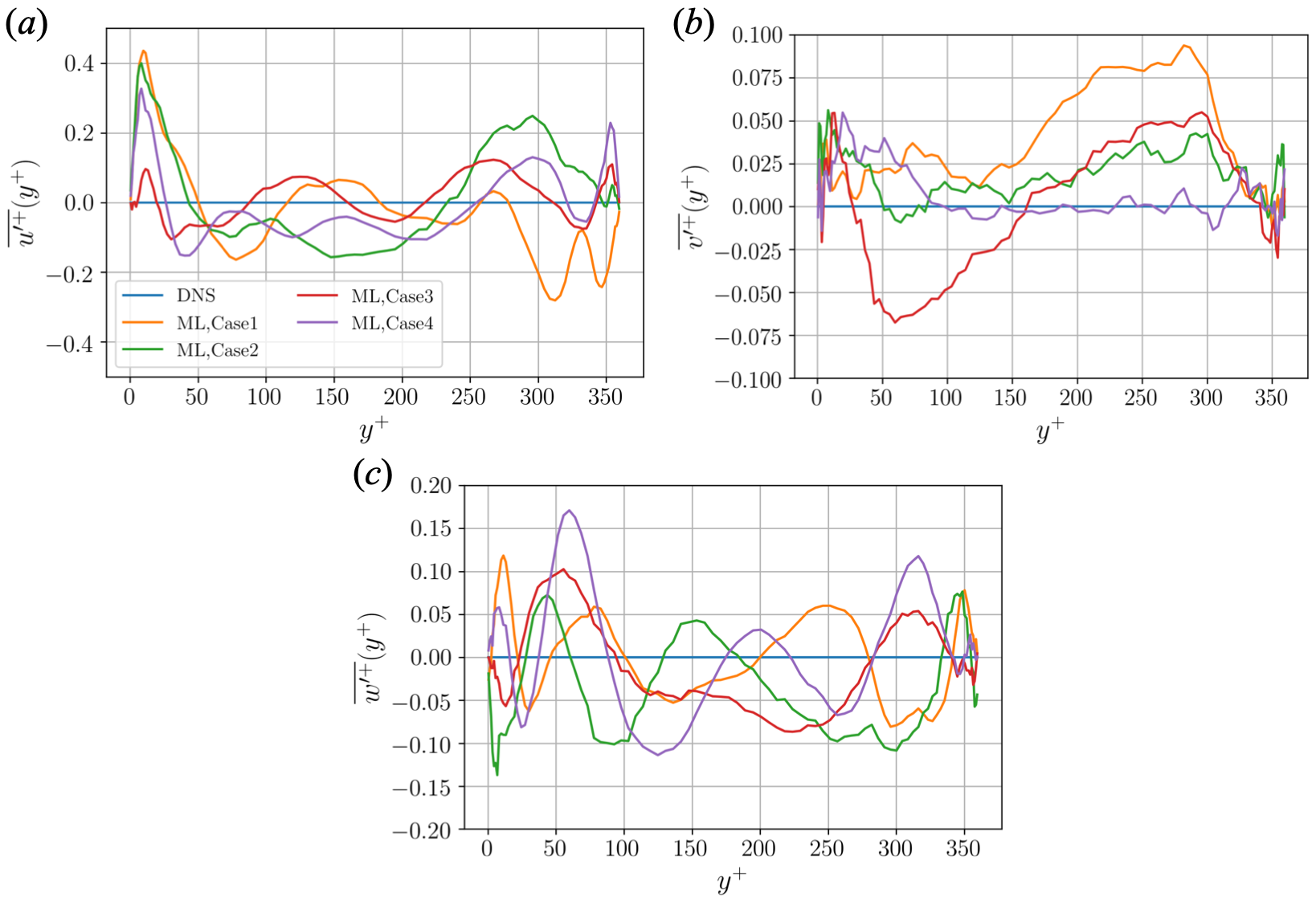}
		\caption{{Mean profiles of ``velocity fluctuations'': ($a$) $\overline{u^\prime}$, ($b$) $\overline{v^\prime}$, ($c$) $\overline{w^\prime}$.}}
		\label{fig5}
	\end{center}
\end{figure}
%%%%%%%%%%%%%%%%%%%%%%%%%%%%%%%%%%%%%%%%%%%%%%%%%%%%%%%

%%%%%%%%%%%%%%%%%%%%%%%%%%%%%%%%%%%%%%%%%%%%%%%%%%%%%%%
%%%%%%TABLE 2-3%%%%%%%%%%%%%%%%%%%%%%%%%%%%%%%%%%%%%%%%%%%%%
%%%%%%%%%%%%%%%%%%%%%%%%%%%%%%%%%%%%%%%%%%%%%%%%%%%%%%%
\begin{table}
\begin{center}
\caption{{Bulk-mean value of ``velocity fluctuations''.}}
\def~{\hphantom{0}}
\begin{tabular}{cccc}
     \hline \hline
            & $\langle\overline{u^{\prime+}}\rangle$ & $\langle\overline{v^{\prime+}}\rangle$ & $\langle\overline{w^{\prime+}}\rangle$\\
            \hline
     DNS    & $-9.67\times10^{-17}$ & $3.63\times10^{-19}$ & $ 1.51\times10^{-19}$\\
     Case 1 & $-6.59\times10^{-3}$  & $3.32\times10^{-2}$  & $ 5.10\times10^{-3}$\\
     Case 2 & $ 5.80\times10^{-2}$  & $2.02\times10^{-2}$  & $-3.45\times10^{-2}$\\
     Case 3 & $ 1.29\times10^{-2}$  & $2.41\times10^{-2}$  & $-7.25\times10^{-2}$\\
     Case 4 & $ 2.40\times10^{-2}$  & $8.27\times10^{-2}$  & $-9.03\times10^{-3}$\\
     \hline \hline
    \end{tabular}
  \label{table23}
\end{center}
\end{table}
%%%%%%%%%%%%%%%%%%%%%%%%%%%%%%%%%%%%%%%%%%%%%%%%%%%%%
%%%%%%%%%%%%%%%%%%%%%%%%%%%%%%%%%%%%%%%%%%%%%%%%%%%%%%

%%% FIGURE 6 %%%%%%%%%%%%%%%%%%%%%%%%%%%%%%%%%%%%%%%%%%%
\begin{figure}
	\begin{center}
		\vspace{0mm}
		\hspace{0mm}
		\includegraphics[width=0.90\textwidth]{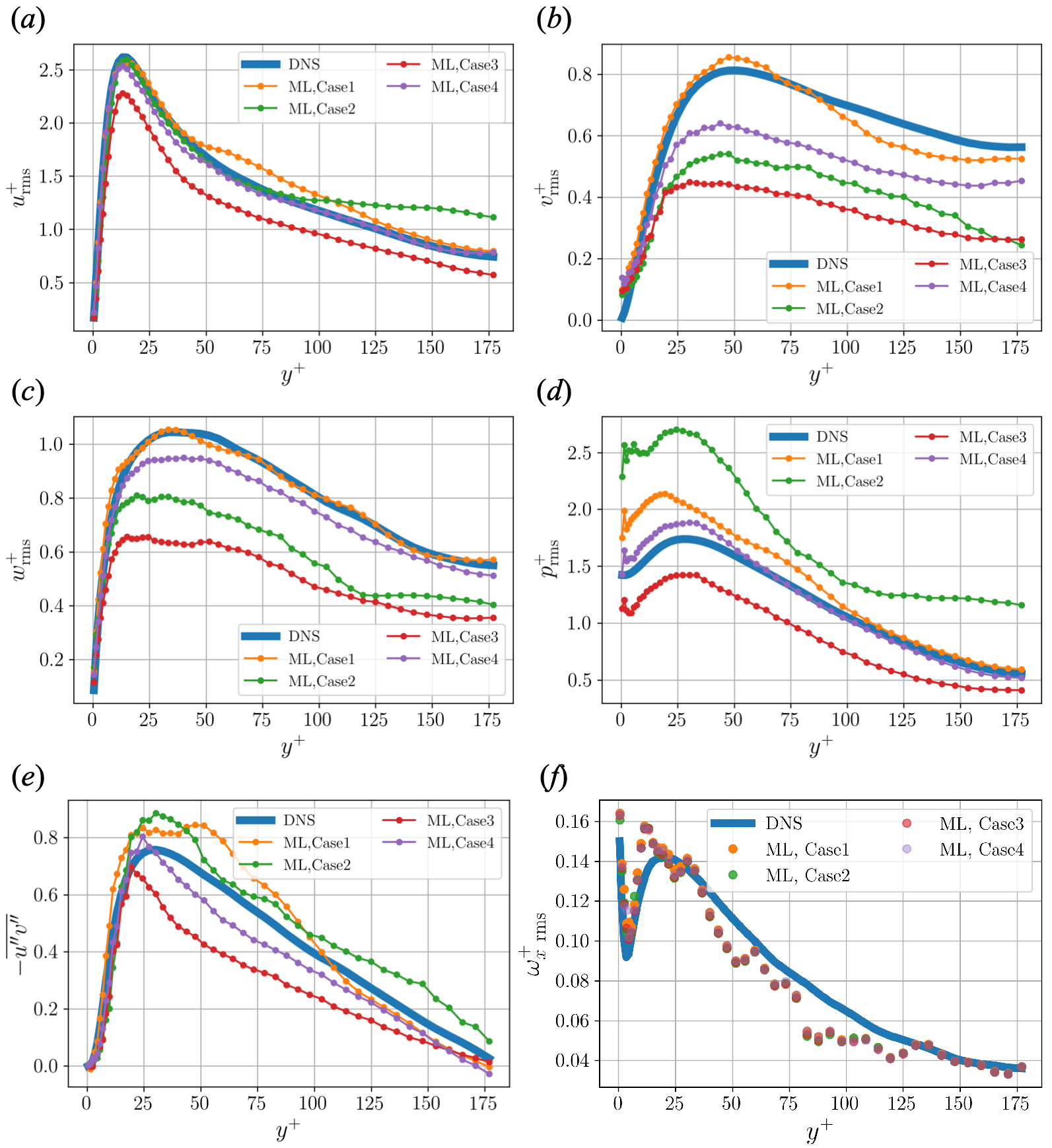}
		\caption{{Statistics based on the corrected fluctuations: ($a$) $u^+_{\rm rms}$, ($b$) $v^+_{\rm rms}$, ($c$) $w^+_{\rm rms}$, ($d$) $p^+_{\rm rms}$, ($e$) $-\overline{u^{\prime\prime+} v^{\prime\prime+}}$, ($f$) $\omega^+_{x~{\rm rms}}$,}}
		\label{fig6}
	\end{center}
\end{figure}
%%%%%%%%%%%%%%%%%%%%%%%%%%%%%%%%%%%%%%%%%%%%%%%%%%%%%%%

%%% FIGURE 7 %%%%%%%%%%%%%%%%%%%%%%%%%%%%%%%%%%%%%%%%%%%
\begin{figure}
	\begin{center}
		\vspace{0mm}
		\hspace{0mm}
		\includegraphics[width=0.70\textwidth]{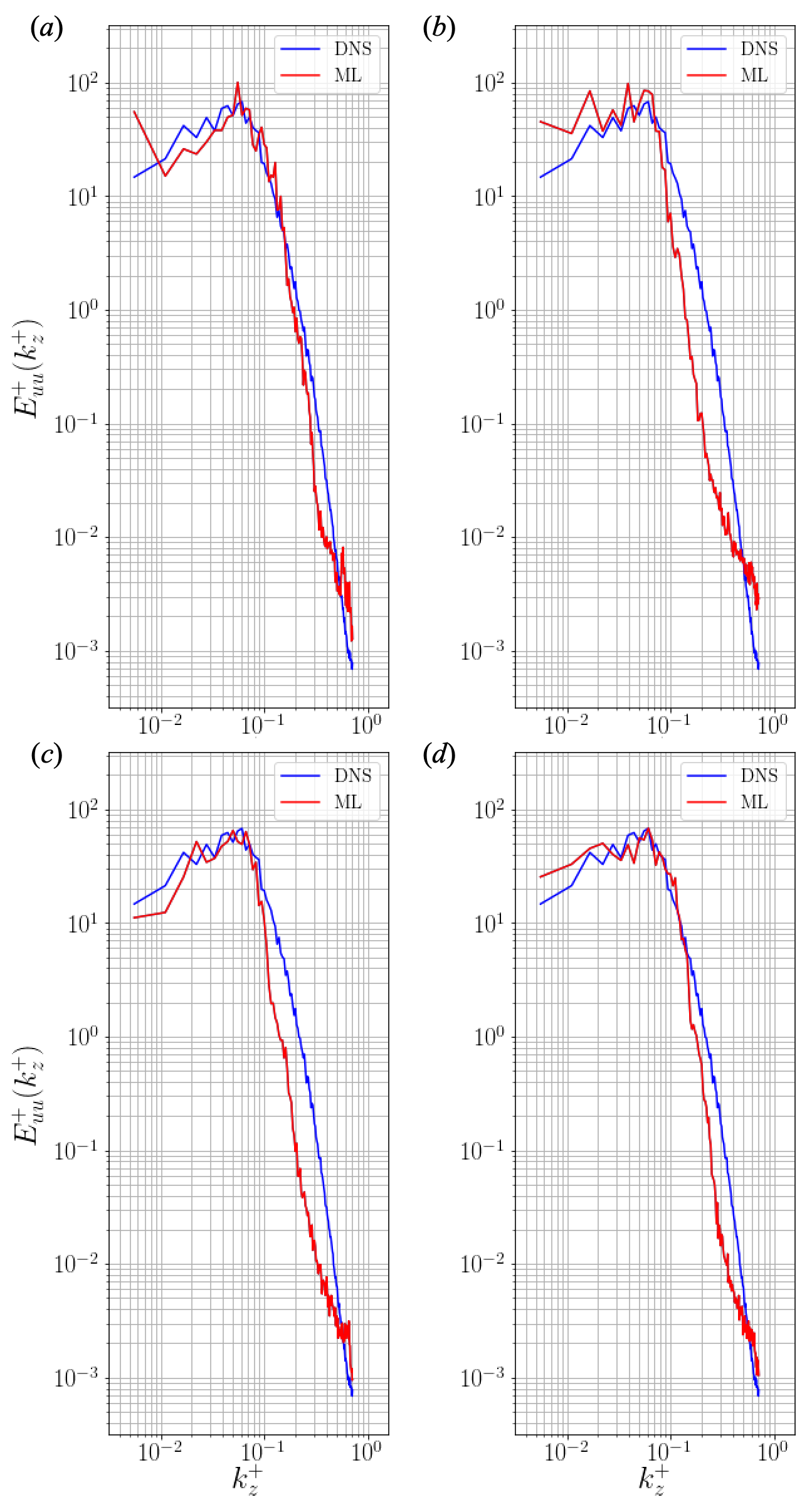}
		\caption{Spanwise energy spectrum: ($a$) Case 1, ($b$) Case 2, ($c$) Case 3, ($d$) Case 4.}
		\label{fig7}
	\end{center}
\end{figure}
%%%%%%%%%%%%%%%%%%%%%%%%%%%%%%%%%%%%%%%%%%%%%%%%%%%%%%%

%%% FIGURE 8 %%%%%%%%%%%%%%%%%%%%%%%%%%%%%%%%%%%%%%%%%%%
\begin{figure}
	\begin{center}
		\vspace{0mm}
		\hspace{0mm}
		\includegraphics[width=0.90\textwidth]{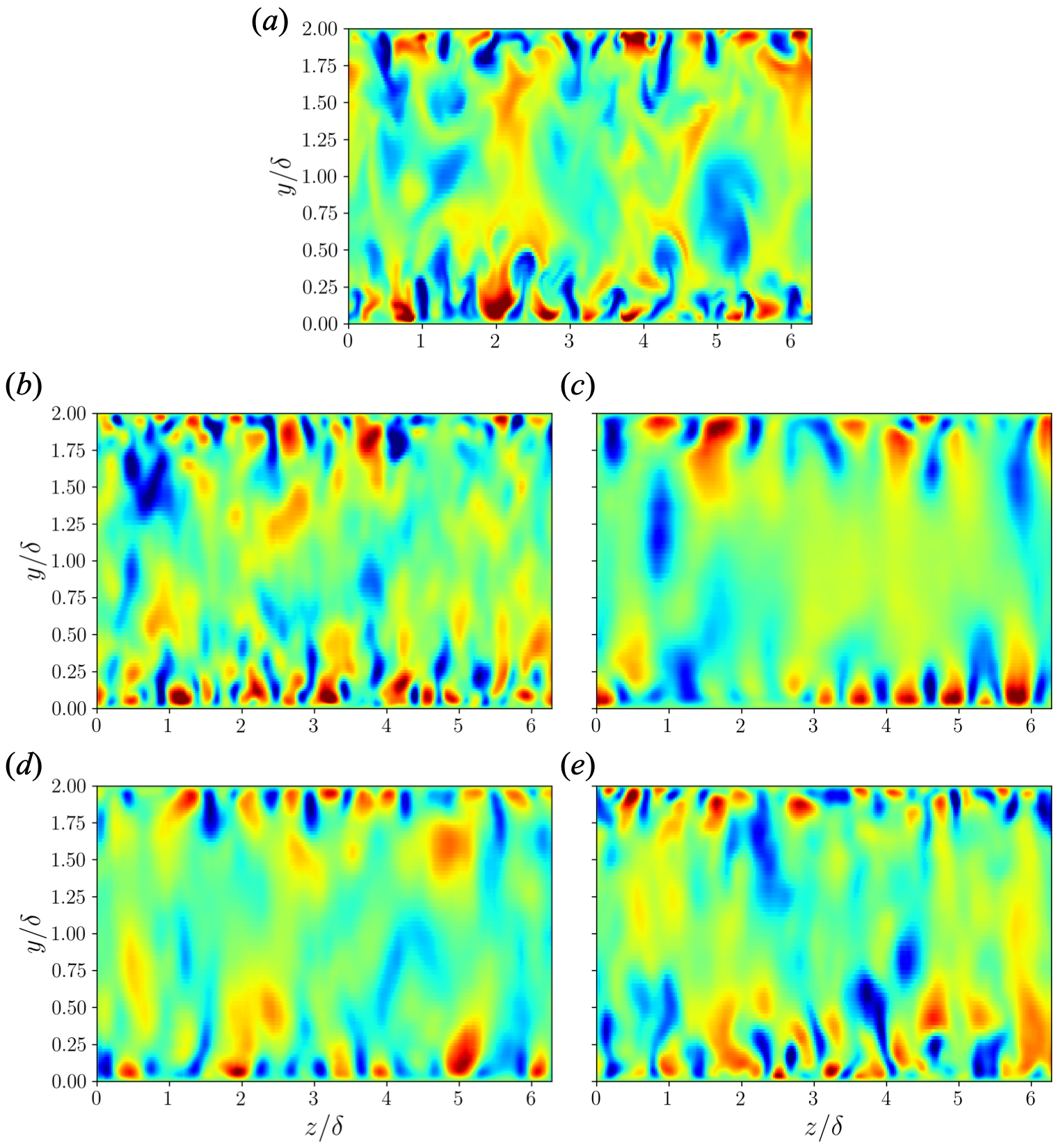}
		\caption{{Cross-sectional contour of streamwise velocity fluctuations $u'$ after recycling for 200 time steps (i.e., about 250 wall unit time): ($a$) DNS, ($b$) Case 1, ($c$) Case 2, ($d$) Case 3, ($e$) Case 4. Animation s are available at: 
		http://kflab.jp/en/index.php?MLTG2.
		}}
		\label{fig8}
	\end{center}
\end{figure}
%%%%%%%%%%%%%%%%%%%%%%%%%%%%%%%%%%%%%%%%%%%%%%%%%%%%%%%

{Figure \ref{fig5} shows the mean profiles of ``velocity fluctuations'', i.e., $\overline{u^{\prime+}}$, $\overline{v^{\prime+}}$, and $\overline{w^{\prime+}}$, where the overbar denotes the average in the spanwise direction and in time. 
Ideally, these quantities should be zero, as indicated by the DNS data shown together. 
However, the distributions obtained by the ML model show substantial non-zero values due to the lack of zero-mean contraint in the learning process. 

The bulk-mean velocities computed from these profiles, $\langle\overline{u^{\prime+}}\rangle$, $\langle\overline{v^{\prime+}}\rangle$, and $\langle\overline{w^{\prime+}}\rangle$, where the brackets denotes the average in wall-normal direction, presented in Table \ref{table23} reveals that the error amounts to 0.5\% of the bulk-mean velocity of the original flow.
In order to fix this problem, we define the corrected fluctuations as
\begin{eqnarray}
{u_i^{\prime\prime}} = {u_i^\prime} - \overline{u_i^\prime} \label{eq7}
\end{eqnarray}
and
\begin{eqnarray}
{p^{\prime\prime}} = {p^\prime} - \overline{p^\prime}.
\end{eqnarray}
The turbulence statistics based on these corrected velocity fluctuations are shown in Fig.~\ref{fig6}. 
Case 1 shows reasonable agreement with the DNS data, while Cases 2-4 show poorer results, especially for $v$ and $w$ components.
}

{To examine the accuracy in the spatial structure reproduced by the machine-learned turbulent inflow generators (MLTG) in greater detail,
the spanwise energy spectrum of the streamwise velocity at $y^+=13.2$ is compared in Fig.~\ref{fig7}.
The machine-learned models show reasonable agreement with the DNS data, although some attenuations are observed in higher wavenumber range.
With a higher compression ratio (i.e., Case 2), the higher wavenumber components are damped more, as observed in figure \ref{fig7}($b$). 
}

The cross-sectional structure of the streamwise velocity fluctuations ($u'$) after 200 time steps {(i.e., about 250 wall unit time)} of recycling within MLTG are shown in figure \ref{fig8}. The temporal evolution is best illustrated by the animation
%(\hyperref[http://kflab.jp/en/index.php?MLTG2]{''http://kflab.jp/en/index.php?MLTG{2}''}).
(http://kflab.jp/en/index.php?MLTG{2})
We can confirm the self-sustaining spatio-temporal evolution similar to that of DNS.
Consistent with the statistics presented above, the structure in Case 1 is observed to be most similar to that of DNS among the present four cases.
%%%%%%%%%%%%%%%%%%%%%%%%%%%%%%%%%%%%%%%%%%%%%%%%%%%%

%%% FIGURE 9 %%%%%%%%%%%%%%%%%%%%%%%%%%%%%%%%%%%%%%%%%%%
\begin{figure}[t!]
	\begin{center}
		\vspace{0mm}
		\hspace{0mm}
		\includegraphics[width=1.00\textwidth]{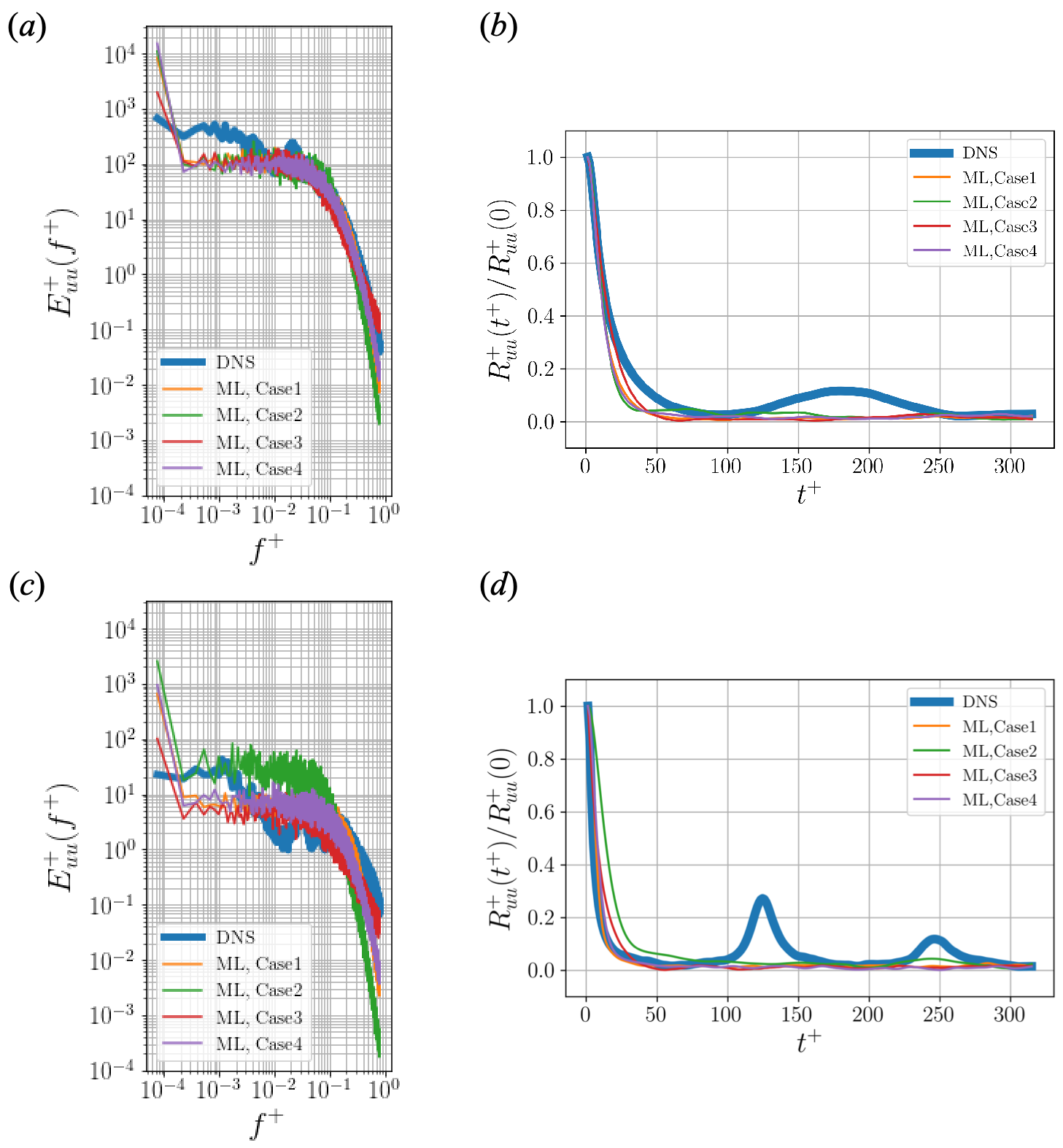}
		\caption{{Temporal statistics in the ${a\;priori}$ test: ($a$) temporal spectrum at $y^+=13.2$; ($b$) temporal two-point correlation coefficient at $y^+=13.2$; ($c$) temporal spectrum at $y^+=177.0$; ($d$) temporal two-point correlation coefficient at $y^+=177.0$.}}
		\label{fig9}
	\end{center}
\end{figure}
%%%%%%%%%%%%%%%%%%%%%%%%%%%%%%%%%%%%%%%%%%%%%%%%%%%%%%%

{
To see whether or not the present MLTG is also suffered from a spurious periodicity issue, we have computed the temporal spectra and two-point correlations of the streamwise velocity component at two different wall-normal locations, i.e., near the wall ($y^+=13.2$) and the channel center ($y^+=177.0$), as shown in Fig.~\ref{fig9}.  
In the case of the driver DNS with a periodic computational domain, we can clearly observe a spurious periodicity with a period corresponding to the length of the computational domain divided by the advection velocity.  
In contrast, with the present MLTG, such a spurious periodicity is not observed.  
This is probably because what the present ML model has learned is not the time sequence of the input data itself but the {\it most probable} nonlinear spatio-temporal relationship between two consecutive time instants.
In other words, the present MLTG is considered to work as a surrogate for the time-discretized nonlinear Navier--Stokes system.}

{The integral time scales, ${\cal T}_u^+$, computed from these temporal two-point correlations, i.e., 
\begin{equation}
{\cal T}_u^+=\int_0^{T^+} \frac{R_{uu}(t^+)}{R_{uu}(0)}dt^+,
\end{equation}
are presented in Table~\ref{timescale}.
The integration time, $T^+$, which should be infinity by definition, is  $T^+=25\;200$ in the present calculation.
%and we took the ensemble-average of XX sets of time sequences.
It can be noticed that the integral time scale in Case 1 is substantially underestimated as compared to that of DNS.
This suggests that, although the spatio-temporal structure is qualitatively well reproduced by the present machine-learned models, the network structure and the parameters need further improvement for more quantitative agreement.  Note that the value near the wall ($y^+=13.2$) in the present DNS, ${\cal T}_u^+\simeq 24$, is also overestimated as compared to that reported in literature, ${\cal T}_u^+\simeq 20$ at $y^+=10$ \citep{Quadrio2003}; this is obviously  due to the insufficient steamwise length and the spurious periodicity thereby.
}

%%%%%%%%%%%%%%%%%%%%%%%%%%%%%%%%%%%%%%%%%%%%%%%
%%%%%%%%%%%%%%%%%%%%%%%%%%%%%%%%%%%%%%%%%%%%%%%
%%%%%%%Table 4%%%%%%%%%%%%%%%%%%%%%%%%%%%%
\begin{table}[t]
\begin{center}
\caption{{The integral time scale, ${\cal T}_u^+$.}}
\def~{\hphantom{0}}
\begin{tabular}{ccc}
     \hline \hline
Location & $y^+=13.2$ \;\;\;& $y^+=177.0$ \\
\hline
     Periodic DNS ($L_x=4\pi\kf{\delta}$)\;\;\; & 24.1 & 15.9 \\
     MLTG Case 1  & 16.7 & 12.3  \\
     MLTG Case 2  & 19.7 & 23.3  \\
     MLTG Case 3  & 19.0 & 13.1  \\
     MLTG Case 4  & 17.8 & 11.8  \\
     \hline \hline
    \end{tabular}
  \label{timescale}
\end{center}
\end{table}
%%%%%%%%%%%%%%%%%%%%%%%%%%%%%%%%%%%%%%%%%%%%%%
%%%%%%%%%%%%%%%%%%%%%%%%%%%%%%%%%%%%%%%%%%%%%
%%%%%%%%%%%%%%%%%%%%%%%%%%%%%%%%%%%%%%%%%%%%%

%%%%%%%%%%%%%%%%%%%%%%%%%%%%%%%%%%%%%%%%%%%%%%%%%%%%
%%%%%%%%%%%%%%%%%%%A posteriori test%%%%%%%%%%%%%%%%%%%%%%%%%
%%%%%%%%%%%%%%%%%%%%%%%%%%%%%%%%%%%%%%%%%%%%%%%%%%%%
%%%%%%%%%%%%%%%%%%%%%%%%%%%%%%%%%%%%%%%%%%%%%%%%%%%%
\subsection{$\bm{A\;posteriori}$ test: inflow-outflow DNS using machine-learned inflow generators}

As an {\it a posteriori} test, we assess whether the machine-learned turbulent inflow generators (MLTG) can actually be used in turbulent flow simulations with inflow condition.
{Following the results of the {\it a priori} test, we use the uncorrected data obtained in Case 1, i.e., $u_i^\prime$ (termed Case 1), and the corrected data, $u_i^{\prime\prime}$ (Case 1$^\prime$), to provide the time-dependent inflow condition for DNS of turbulent channel flow with inflow-outflow condition,} and compare with the results computed using the traditional driver DNS (i.e., an additional periodic DNS as a driver).
\kf{The streamwise length of the compuational domain for the inflow-outflow DNS is $L_x=4\pi\delta$. The convective outflow condition is imposed at the outlet.}
The cross-sectional velocity field data by MLTG is generated every 10 time step of DNS, i.e., $\Delta t^+ = 1.26$.
The inflow data at intermediate time instants are given by using a linear interpolation.

%%% FIGURE 10 %%%%%%%%%%%%%%%%%%%%%%%%%%%%%%%%%%%%%%%%%%%
\begin{figure}[t!]
	\begin{center}
		\vspace{0mm}
		\hspace{0mm}
		\includegraphics[width=0.80\textwidth]{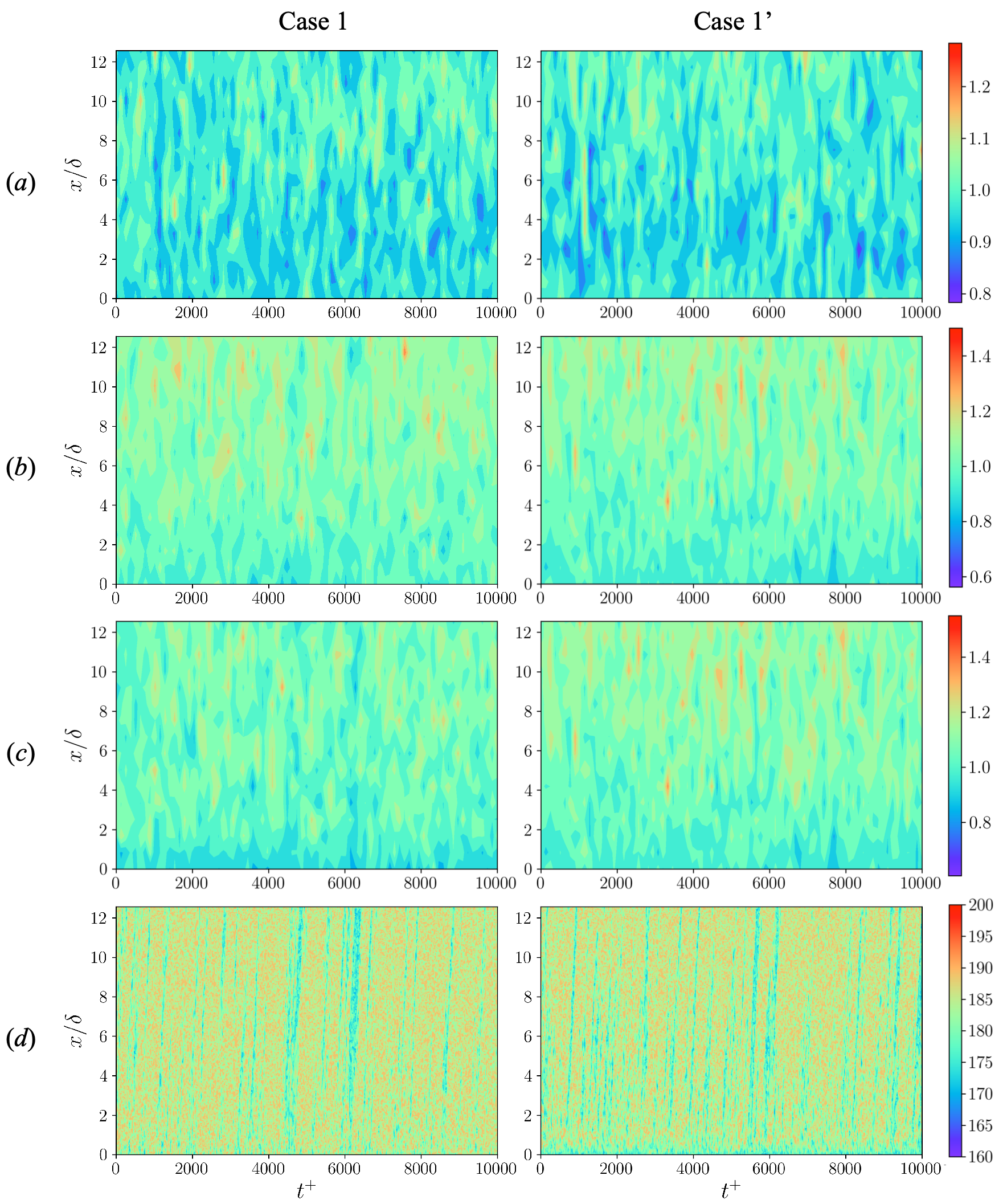}
		\caption{Spatio-temporal development of peak RMS velocity: ($a$) $u'_{\rm MLTG}/u'_{\rm DNS}$; ($b$) $v'_{\rm MLTG}/v'_{\rm DNS}$; ($c$) $w'_{\rm MLTG}/w'_{\rm DNS}$; ($d$) ${\rm Re}_{\tau}$.}
		\label{fig10}
	\end{center}
\end{figure}
%%%%%%%%%%%%%%%%%%%%%%%%%%%%%%%%%%%%%%%%%%%%%%%%%%%%%%%

The spatio-temporal development of the peak value in the RMS velocity fluctuations normalized by the value of DNS, $u'_{i,{\rm MLTG}}/u'_{i,{\rm DNS}}$, and the spatio-temporal development of local friction Reynolds number, $\rm Re_{\tau}$, computed by DNS with MLTG are shown in Fig.~\ref{fig10}.
{Hereafter, the velocity fluctuations in Case 1$^\prime$ are also denoted by a single prime for notational simplicity.}
The horizontal axis is the wall unit time $t^+$ and the vertical axis represents the streamwise length from the inlet.
Note that the peak values of RMS velocities are computed in 16 subsections divided in the streamwise direction.
The computations used the present machine-learned turbulence generator are continued at $t^+= 10\;000$ which is considered long enough to accumulate turbulent statistics, and
$u'_{i,{\rm MLTG}}/u'_{i,{\rm DNS}}$ and $\rm Re_{\tau}$ are maintained {nearly constant}.
\kf{Although the velocity fluctuations are slightly damped near the inlet due the errors of the MLTG, the flow recovers to have the correct statistics after $x/\delta \simeq 2$.}

%%% FIGURE 11 %%%%%%%%%%%%%%%%%%%%%%%%%%%%%%%%%%%%%%%%%%%
\begin{figure}
	\begin{center}
		\vspace{0mm}
		\hspace{0mm}
		\includegraphics[width=0.90\textwidth]{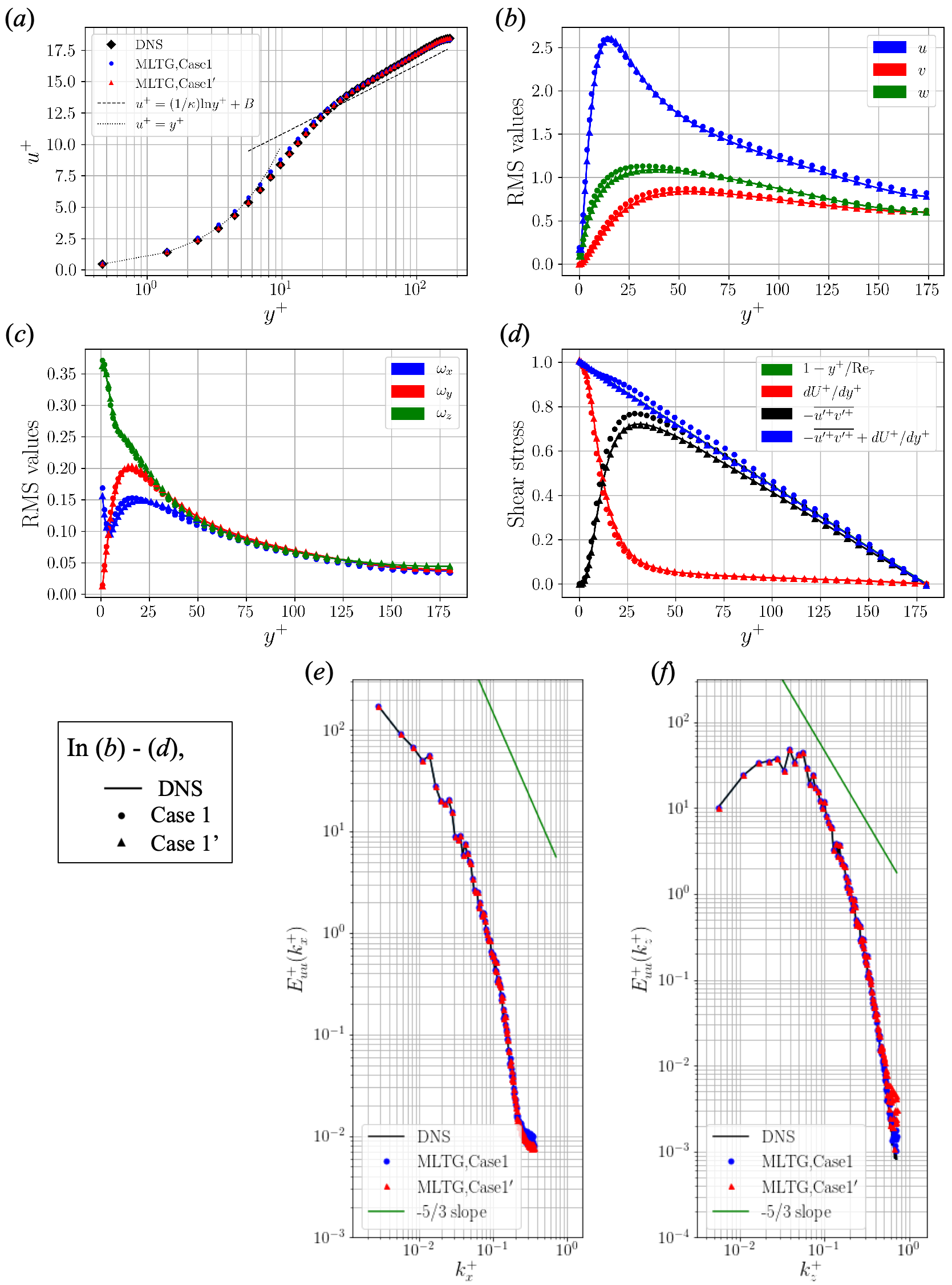}
		\caption{Turbulence statistics in $\sl a$ $\sl posteriori$ test using machine-learned turbulence generator: ($a$) mean velocity profile; ($b$) RMS of $u'_i$; ($c$) RMS of $\omega_i'$; ($d$) shear stress balance; ($e$) streamwise energy spectrum of $u'$; ($f$) spanwise energy spectrum of $u'$.}
		\label{fig11}
	\end{center}
\end{figure}
%%%%%%%%%%%%%%%%%%%%%%%%%%%%%%%%%%%%%%%%%%%%%%%%%%%%%%%

Turbulence statistics computed in the inflow-outflow DNS with the \kf{MLTG} are shown in \kf{Fig.}~\ref{fig11}\kf{. 
The turbulence statistics here are accumulated in the entire computational domain of $L_x=4\pi\delta$}
and normalized in the wall unit of the origianl flow, i.e., ${\rm Re}_\tau=180$.
\kf{The statistics obtained in} the DNS in a periodic domain of $L_x=4\pi\kf{\delta}$ \kf{are also shown for comparison.} 
Mean velocity profile, RMS velocity fluctuations, RMS vorticity fluctuations, Reynolds shear stress, and streamwise and spanwise spectra of streamwise velocity are all in reasonable agreement with the periodic DNS, which confirms that the present machine-learned model properly works as the inflow generator.
{In particular, Case $1^\prime$ based on the corrected fluctuations outperforms Case 1.
The deviation in Case 1 is attibuted to the increased flow rate due to the non-zero mean component as observed in the $\it a\;priori$ test.}

\kf{The present result demonstrates that the turbulence statistics can be reproduced in an inflow-outflow DNS with an MLTG as far as the MLTG reproduces the spatio-temporal charactaristics of inflow turbulence reasonably well, although not perfect.
Even if a small amount of error is contained at the inlet, the flow recovers in the DNS domain to have the correct statistics.
More extensive work, however, is needed to clarify how much and what kind of errors are allowed at the inlet.
}

%%%%%%%%%%%%%%%%%%%%%%%%%%%%%%%%%%%%%%%%%%%%%%%%%%%%%%%
%%%%%%TABLE 4%%%%%%%%%%%%%%%%%%%%%%%%%%%%%%%%%%%%%%%%%%%%%
%%%%%%%%%%%%%%%%%%%%%%%%%%%%%%%%%%%%%%%%%%%%%%%%%%%%%%%
\begin{table}
\begin{center}
\caption{Comparison of computation time in $\sl a$ $\sl posteriori$ test}
\def~{\hphantom{0}}
\begin{tabular}{cccccc}
     \hline \hline
     Generator type & Time (s) & Ratio versus & Ratio versus \\
                    &          & MLTG, Case 1 (CPU) & MLTG, Case 1 (GPU) \\ \hline
     Driver DNS ($L_x=2\pi\kf{\delta}$) & 2.39 & 181 & 582\\
     MLTG, Case 1 (CPU) & $1.32\times10^{-2}$&1.00&3.21 \\
     MLTG, Case 1 (GPU) & $4.11\times10^{-3}$&0.311&1.00 \\
     \hline \hline
    \end{tabular}
  \label{table4}
\end{center}
\end{table}
%%%%%%%%%%%%%%%%%%%%%%%%%%%%%%%%%%%%%%%%%%%%%%%%%%%%%
%%%%%%%%%%%%%%%%%%%%%%%%%%%%%%%%%%%%%%%%%%%%%%%%%%%%%%

At last, the computational time required for generating the turbulent inflow data for one time step is compared shown in table \ref{table4}.
When we use a periodic DNS as a driver simulation, at least $L_x=2\pi\kf{\delta}$ should be required to obtain reasonable statistics.
Therefore, comparison is made between the present model and a periodic DNS with $L_x=2\pi\kf{\delta}$.
Although the concrete value of computational time highly depends on the environment such as the machine, complier, and library used and the way of coding, the MLTG is apparently faster than the driver-type turbulence generator. 
Under our environment, the computational speed of the machine-learned model for generating one cross-sectional velocity field is about 180 times faster on CPU (single core of Intel Xeon E5-2680v4, 2.4~GHz) and about 580 times faster when a GPU (NVIDIA Tesla K40) is used than the driver DNS run on the same CPU.
In addition, considering the fact that in the present machine-learned case computes a cross-sectional velocity field every 10 time step of DNS (since MLTG is not restricted by the Courant number), the actual speed-up rate is 10 times the values above; namely, 1\;800 times and { {5\;800}} times faster when CPU and GPU are used, respectively.

In sum, it can be concluded that the present type of MLTG can be used also in practical simulations in terms of self-sustainability of turbulent structure, accuracy in reproduced turbulent statistics, and low computational cost.

%%%%%%%%%%%%%%%%%%%%%%%%%%%%%%%%%%%%%%%%%
%%%%%%%%%%%%%%%%%%%%%%%%%%%%%%%%%%%%%%%%%
%%%%%%%%%%%%%%%%CONCLUSION%%%%%%%%%%%%%%%%%%
%%%%%%%%%%%%%%%%%%%%%%%%%%%%%%%%%%%%%%%%%%
%%%%%%%%%%%%%%%%%%%%%%%%%%%%%%%%%%%%%%%%%%
%%%%%%%%%%%%%%%%%%%%%%%%%%%%%%%%%%%%%%%%%%
\section{Conclusions}
In this paper, we proposed a machine-learned turbulence generator (MLTG) using {an autoencoder-type convolutional neural network (CNN) combined with a multi-layer perceptron (MLP)}.
For the test case, a turbulent channel flow at the friction Reynolds number of ${\rm Re}_{\tau} = 180$ is considered as a first step.

The machine-learn{ing (ML)} models were trained using a series of instantaneous velocity fields in a single cross-section obtained by direct numerical simulation (DNS) so as to output the cross-sectional velocity field at a specified future time instant.
In {the} {\it a priori} test, the present {MLTG} was found to accurately reproduce not only the turbulence statistics but also the spatio-temporal development of cross-sectional structure{, although some deviation in the flow rate was found}.
{Moreover, unlike the conventional driver DNS using a periodic domain, the present {MLTG} is found to be free from the spurious periodicity.} 
{A}s an {\it a posteriori} test, we performed DNS of inflow-outflow turbulent channel flow with the trained MLTG as the time-dependent inflow condition.
The MLTG was able to maintain the turbulent channel flow in a long time period up to $10\;000$ wall unit time, which is sufficient to accumulate turbulent statistics, with much lower computational cost than the conventional driver simulation.

The present results suggest that MLTG is an attractive alternative to the conventional methods.
{Although there is a computational overhead for training, MLTG should be useful in the cases where many simulations are performed under statistically the same inflow condition but different downstream conditions due to, e.g., control, roughness, and obstacles.}
Extension of the proposed methodology to other types of flows, such as spatially developing boundary layer and flows around a body, is straightforward, but the accuracy should be assessed for each problem. 

{From the observation of the $\it a~priori$ test, the results are found to be sensitive against the parameters of machine learning including the number of layers, units and so on.
Although we have obtained reasonable results in the present study, more extensive study should be made to find better network structures giving higher accuracy.
For instance, 
the long short term memory (LSTM) \citep{HS1997} proposed to deal with the complicated time series of data can be considered to increase the accuracy in temporal characteristics ---
in fact, usefulness of LSTM has recently been demonstrated by \citet{Vlachas2018} for a number of dynamical systems. 
In addition, an architecture independent of the shape and size of input data will also be needed.
Also, the structure of CNN can be modified so as to learn the different spatial scales more accurately, which is one of the ongoing studies in our group \citep{Fukami2018}.

Despite the merits mentioned above, the major drawback of the present method in contrast to the conventional synthetic turbulence generators is that not only the statistics but also spatio-temporal data of the target flow are still required to train the network.
The ultimate goal may be to construct a similar network which requires lower order information such as spatio-temporal correlations only.}
However, {we believe that} the present study {will} serve as a {good} starting point {toward} this direction {---} the remaining issues will be tackled in the future.

\section*{Acknowlegments}
{Authors} are grateful to Dr. S. Obi, Dr. K. Ando, Dr. Y. Aoki, and Mr. K. Endo (Keio Univ.), Dr. M. Yamamoto and Dr. T. Tsukahara (Tokyo Univ. of Science), Dr. K. Iwamoto (Tokyo Univ. of Agriculture and Technology), Dr. Y. Hasegawa (The Univ. of Tokyo), Dr. N. Fukushima (Tokai Univ.), Dr. H. Mamori (Univ. of Electro-Communications), and Dr. P. Koumoutsakos (ETH Zurich) for fruitful discussion, {for which K. Fukami and K. Fukagata also thank Dr. K. Taira (UCLA).}
This work was supported through JSPS KAKENHI Grant Number 18H03758 by Japan Society for the Promotion of Science (JSPS).

%%%%%%%%%%%%%%%%%%%%%%%%%%%%%%%%
%%%%%%%%%%%%%%%%%%%%%%%%%%%%%%%%%
%%%%%%%%%%%%%%%%%%%%%%%%%%%%%%%%%

% Create the reference section using BibTeX:
%\bibliography{your-bib-file}

\end{document}